
%
%
%
\input phyzzx
\tolerance=5000
\sequentialequations
\def\rl{\rightline}

\def\r#1{$\bf#1$}

\def\t1{{\tilde 1}}

\def\AEF{A.E. Faraggi}
\def\DVN{D. V. Nanopoulos}

\def\NPB#1#2#3{Nucl. Phys. B {\bf#1} (19#2) #3}
\def\PLB#1#2#3{Phys. Lett. B {\bf#1} (19#2) #3}
\def\PRD#1#2#3{Phys. Rev. D {\bf#1} (19#2) #3}

\def\PRT#1#2#3{Phys. Rep. {\bf#1} (19#2) #3}

\def\IJMP#1#2#3{Int. J. Mod. Phys. A {\bf#1} (19#2) #3}

\def\vev#1{\left\langle #1\right\rangle}
\def\l{\langle}
\def\r{\rangle}

\REF\GSW{For a review see, M. Green, J. Schwarz and E. Witten,
Superstring Theory, 2 vols., Cambridge
University Press, 1987.}
\REF\SUSY{For reviews see,
H.P. Nilles, \PRT{110}{84}{1}; \DVN~ and Lahanas, \PRT{145}{87}{1}.}
\REF\Louis{For a review see, J. Louis, SLAC-PUB-5645, DPF Conference, 1991.}
\REF\FFM{I. Antoniadis, J. Ellis, J. Hagelin, and \DVN, \PLB{231}{89}{65};
\AEF, D.V. Nanopoulos and K. Yuan, \NPB{335}{90}{347};
I. Antoniadis, G. K. Leontaris and J. Rizos, \PLB{245}{90}{161};
J. Lopez, D.V. Nanopoulos and K. Yuan, \NPB{399}{93}{654}, hep-th/9203025.}
\REF\SRSM{M. Dine {\it{el al.}}, \NPB{259}{85}{549};
B. Greene, K.H. Kirklin, P.J. Miron and G.G. Ross, \NPB{292}{87}{606};
A. Font, L.E. Ibanez, F. Quevedo and A. Sierra, \NPB{331}{90}{421};
D. Bailin, A. Love and S. Thomas,
\PLB{194}{87}{385}.}
\REF\EU{\AEF, \PLB{278}{92}{131}.}
\REF\TOP{\AEF, \PLB{274}{92}{47}.}
\REF\SLM{\AEF, \PRD{47}{93}{5021}; \NPB{387}{92}{239}, hep-th/9208024.}
\REF\FFF{I. Antoniadis, C. Bachas, and C. Kounnas, \NPB{289}{87}{87};
I. Antoniadis and C. Bachas, \NPB{298}{88}{586};
H. Kawai, D.C. Lewellen, and S.H.-H. Tye,
Nucl. Phys. B {\bf 288} (1987) 1; R. Bluhm, L. Dolan, and P. Goddard,
Nucl. Phys. B {\bf 309} (1988) 330.}
\REF\NRT{A.E. Faraggi, \NPB{403}{93}{101}, hep-th/9208023; IASSNS--HEP--94/18,
hep-ph/9403312.}
\REF\GCU{\AEF, \PLB{302}{93}{202}, hep-ph/9301268.}
\REF\FM{A.E. Faraggi, \NPB{407}{93}{57}, hep-ph/9210256.}
\REF\NM{\AEF~ and E. Halyo, \PLB{307}{93}{311}, hep-th/9303060.}
\REF\CKM{\AEF~ and E. Halyo, \PLB{307}{93}{305}, hep-ph/9301261;
	 \AEF~ and E. Halyo, \NPB{416}{94}{63}, hep-ph/9306235.}
\REF\HALYOI{E. Halyo, \PLB{318}{93}{597}, hep-ph/9308211.}
\REF\FOC{\AEF, IASSNS--HEP--93/67, Phys. Lett. B, in press, hep-ph/9311312.}
\REF\DSW{M. Dine, N. Seiberg and E. Witten, \NPB{289}{87}{589}.}
\REF\ADS{J.J. Atick, L.J. Dixon and A. Sen, \NPB{292}{87}{109};
	 S. Cecotti, S. Ferrara and M. Villasante, \IJMP{2}{87}{1839}.}
\REF\KLN{S. Kalara, J. Lopez and D.V. Nanopoulos,
\PLB{245}{91}{421}; \NPB{353}{91}{650}.}
\REF\AEFNT{I. Antoniadis, J. Ellis, E. Floratos,
	   \DVN ~and T. Tomaras, \PLB{191}{87}{96}.}
\REF\FGKP{S. Ferrara, L. Girardello, C. Kounnas, M. Porrati,
\PLB{192}{87}{368}; \PLB{194}{87}{358}.}
\REF\GPR{For a recent review see,
A. Giveon, M. Porrati and E. Rabinovici, RI--1--94, NYU--TH.94/01/01,
hepth/9401139.}
\REF\DKL{L. Dixon, V. Kaplunovsky and Louis, \NPB{329}{90}{27};
\NPB{355}{91}{649}.}
\REF\KLNII{S. Kalara, J.L. Lopez, \DVN, \PLB{287}{92}{82}, hep-ph/9112011.}
\REF\SQCD{D. Amati {\it et. al.}, \PRT{162}{88}{169}.}
\REF\LT{D. L\"ust and T. Taylor, \PLB{253}{91}{335}.}
\REF\AELN{I. Antoniadis, J. Ellis, A.B. Lahanas, \DVN, \PLB{241}{90}{24}.}
\REF\KLNI{S. Kalara, J.L. Lopez, \DVN, \PLB{275}{91}{304}, hep-ph/9110023.}
\REF\ROSS{J.A. Casas, Z. Lalak, C. Munoz, G.G. Ross, \NPB{347}{90}{243}.}
\REF\CVETIC{M. Cvetic {\it et. al.}, \NPB{361}{91}{194}.}
\REF\S{N. Seiberg, RU--94--18, hepth/9402044.}
\REF\DiS{N.V. Krasnikov, \PLB{193}{87}{37}; L. Dixon, in Proc. A.P.S.
DPF Meeting, Houston, TX. 1990; T. Taylor, \PLB{252}{90}{59}.}
\REF\GSUSYB{M. Dine, R. Rohm, N. Seiberg and E. Witten, \PLB{156}{85}{55}.}
\REF\KL{V. Kaplunovsky and J. Louis, \PLB{306}{93}{269},
        hepth/9303040;
        R. Barbieri, J. Louis and M. Moretti, \PLB{312}{93}{451},
        hepph/9305262.}
\REF\IL{L. Ibanez and D. L\"ust, \NPB{382}{92}{305}, hepth/9202046.}

\singlespace
\rl{IASSNS--HEP--94/17}
\rl{WIS--94/13/MAR--PH}
\rl{\today}
\rl{T}
\pagenumber=1
\normalspace
\smallskip
\titlestyle{\bf{Hierarchical Supersymmetry Breaking in Superstring Derived
Standard--like Models}}
\smallskip
\centerline{Alon E. Faraggi{\footnote*{e--mail address: faraggi@sns.ias.edu}}}
\smallskip
\centerline {School of Natural Sciences, Institute of Advanced Studies}
\centerline {Princeton, NJ 08540}
\smallskip
\centerline {and}
\smallskip
\centerline{Edi Halyo
{\footnote\dag{e--mail address: jphalyo@weizmann.bitnet}}}
\smallskip
\centerline {Department of Particle Physics, Weizmann Institute of Science}
\centerline {Rehovot 76100, Israel}
\smallskip

\titlestyle{\bf ABSTRACT}

We examine the problem of supersymmetry breaking in realistic superstring
standard--like models which are constructed in the free fermionic
formulation. We impose a supersymmetric vacuum at the Planck
scale by requiring vanishing F and D constraints at the cubic level of the
superpotential. We then study possible scenarios for supersymmetry
breaking by examining the role of nonrenormalizable terms and
hidden sector gaugino and matter condensates.
We argue that in some scenarios hierarchical
supersymmetry breaking in the observable sector is possible.

\singlespace
\vskip 0.5cm
\nopagenumbers
\pageno=0
\endpage
\normalspace
\pagenumbers

\centerline{\bf 1. Introduction}

Experiments at present day accelerators indicate that the Standard Model
correctly accounts for the observed particle spectrum and interactions.
Due to its renormalizability, the Standard Model
may be the correct theory up to a very large energy scale.
Indeed, for the last two decades this possibility has been exploited in
the development of Grand Unified Theories  and superstring theories [\GSW].
However, in this context, two fundamental problems are raised: the
stabilization of the scalar sector and the derivation of the low energy scale
from the fundamental high energy scale.
Supersymmetry and supergravity
[\SUSY] have been advocated as possible solutions
to both of these problems. However, as
supersymmetry is not a symmetry of the observed particle spectrum and
interactions, it is at best a broken symmetry.
The problem of supersymmetry breaking received much attention both in the
context of supergravity and superstring theories [\Louis]. In
supergravity, one has to impose the existence of a hidden sector.
Strong hidden
sector dynamics are responsible for breaking supersymmetry in the
hidden sector, which is transmitted to the observable sector
by gravitational interactions. In the context of superstring theories, the
hidden sector arises naturally from the consistency of the theory.

On the other hand, superstring theory is the only known candidate for a
consistent unified theory of quantum gravity and the gauge interactions.
As such it provides a unique opportunity to study how the parameters of the
Standard Model may arise from a fundamental Planck scale theory. Luckily,
to produce a fermion spectrum string theory must accommodate $N=1$ space--time
supersymmetry, at least at the Planck scale. Several attempts have been
made to construct semi--realistic superstring models [\FFM--\SLM]. Of those,
the most notable are the realistic models [\FFM,\EU,\TOP,\SLM]
in the free fermionic formulation [\FFF].
Due to proton decay constraints, the most compelling choice
is to derive the Standard Model directly from the superstring [\SLM,\NRT].
The superstring derived
standard--like models of Refs. [\EU,\TOP,\SLM] exemplify how
the Standard Model may be derived directly from superstring theory.
In Refs. [\SLM,\NRT,\GCU,\FM,\NM,\CKM,\HALYOI]
some of the fundamental problems in particle physics
were addressed in the context of these models. Among them, proton decay
[\SLM,\NRT], gauge coupling unification [\GCU], generation mass
hierarchy [\FM], neutrino masses [\NM],
texture of fermion mass matrices [\NRT,\FM,\CKM], quark flavor mixing [\CKM],
and axions [\HALYOI].

In this paper, we study the problem of supersymmetry breaking in realistic
superstring derived standard--like models.
These models correspond to $Z_2\times Z_2$
orbifold models with nontrivial background fields and with three additional
Wilson lines [\FOC].
Three additional Wilson lines are needed to reduce the number
of generations to three, one from each one of the twisted sectors
of the corresponding
$Z_2\times Z_2$ orbifold models. The standard--like models contain a
hidden gauge group which is some subgroup of $E_8$, typically
$SU(5)\times SU(3)\times U(1)^2$, with matter spectrum in vector--like
representations. An important property of the standard--like models
is the existence of an ``anomalous'' $U(1)$ symmetry,
which is instrumental in determining the parameters of the low energy
effective theory and in lifting the degeneracy among string vacua.

The problem of supersymmetry breaking is divided into two parts.
The first is the determination of the compactification parameters,
{\it i.e.} the dilaton VEV, the moduli VEVs and the singlet VEVs which cancel
the anomalous $U(1)$ D--term equation. The second is the hierarchical
breaking of supersymmetry in the observable sector,  given a supersymmetric
vacuum at the Planck scale.
In this paper, we do not address the first part
of the problem. We assume that these VEVs are fixed by some unknown,
possibly nonperturbative,
string mechanism. Thus, given a set of compactification
parameters with a supersymmetric vacuum at the Planck scale and at the
cubic level of the superpotential, we examine a possible scenario
for supersymmetry breaking at hierarchically low energies. In our approach,
we impose a supersymmetric vacuum
at the Planck scale by requiring vanishing F and D flatness constraints
at the cubic level of the superpotential.
The $SO(10)$ singlet VEVs that we impose are motivated
by requiring quark masses and mixing of the correct order of magnitude.
By imposing vanishing F and D flat directions at the cubic level
of the superpotential we guarantee
that there are no supersymmetry breaking terms of the
order of $M_{Pl}$.
We then study the effects of strong hidden sector dynamics and
nonrenormalizable terms on the effective superpotential. We find that
if only  observable sector states obtain nonvanishing VEVs in
the cancelation of the ``anomalous'' $U(1)$ D--term equation,
supersymmetry is unbroken to all orders of nonrenormalizable terms.
On the other hand, if some hidden sector states obtain
nonvanishing VEVs, then supersymmetry
is broken due to nonrenormalizable terms and
hidden sector matter condensates. When hidden sector matter condensation
occurs,
the cubic level F constraints are modified to a set of equations that
cannot be satisfied simultaneously. Consequently,
nonvanishing F--terms are generated which break supersymmetry at
a hierarchically small scale in the observable sector.

Our paper is organized as follows. In section 2, we review the
superstring standard--like models and the properties that are important
for the discussion that follows. In section 3, we discuss the cubic level
superpotential and phenomenological constraints on the Standard Model
singlet VEVs.
In section 4, we examine the effects of
nonrenormalizable terms and the strong hidden sector dynamics.
Section 5 contains a discussion of several problems related to
supersymmetry breaking and our conclusions.

\bigskip
\centerline{\bf 2. The superstring standard--like models}

The superstring standard--like models [\TOP,\EU,\SLM]
are constructed in the four dimensional free fermionic formulation [\FFF].
To study the problem of supersymmetry breaking we focus on the model that was
presented in Ref. [\EU].
The standard--like models are generated
by sets of eight basis vectors,
$\{{\bf 1},S,b_1,b_2,b_3,\alpha,\beta,\gamma\}$. The set
$\{{\bf 1},S,b_1,b_2,b_3,2\gamma\}$ is common to all the realistic
models in the free fermionic formulation. The set
$\{{\bf 1},S,{\bf1}+b_1+b_2+b_3,2\gamma\}$ generates a toroidal compactified
model with $N=4$ space--time supersymmetry and
$SO(12)\times SO(16)\times SO(16)$ gauge symmetry. The vectors $b_1$ and
$b_2$ correspond to moding out the six dimensional torus by a $Z_2\times Z_2$
discrete symmetry with standard embedding, [\FM,\FOC].
The vectors $\alpha$, $\beta$, $\gamma$ differ
between models and correspond to different choices of Wilson lines in the
orbifold language. The various choices of vectors $\alpha$, $\beta$, $\gamma$
and of the phases $c\left(\matrix{\alpha,\beta,\gamma\cr
				   {\bf1},S,b_i\cr}\right)$
fix the physical spectrum and determine the low energy effective theory
of the superstring standard--like models.

The full massless spectrum together with the quantum numbers were
given in Ref. [\EU].
The gauge group after all GSO projections have been applied
is{\footnote*{$U(1)_C={1\over2}U(1)_{B-L}$,
	 $U(1)_L={1\over2}U(1)_{T_{3_R}}$.}}
$\{SU(3)\times SU(2)\times U(1)_C\times U(1)_L\times U(1)^6\}_{_O}
\times\{SU(5)_H\times SU(3)_H\times U(1)^2\}_{_H}$,
where the first curly brackets
correspond to the observable gauge group that arises from the first $SO(16)$
times $SO(12)$. The second curly brackets correspond to the hidden gauge
group that arises from the second $SO(16)$.
Here by hidden gauge group we mean that the states that are identified
with the three generations do not transform under the hidden gauge group.
The weak hypercharge is uniquely given by
$U(1)_Y={1\over3}U(1)_C+{1\over2}U(1)_L$. The orthogonal combination
is given by $U(1)_{Z^\prime}=U(1)_C-U(1)_L$.
The supersymmetry generator is the basis vector $S$ and the superpartners of
the states from a given sector $\alpha$ are obtained from the sector
$S+\alpha$. The sectors $b_1$, $b_2$ and $b_3$ correspond to
the three twisted sectors of the orbifold model and produce three
16 of $SO(10)$ decomposed under
$SU(3)\times SU(2)\times U(1)_C\times U(1)_L$ with charges
under the horizontal symmetries.
For every generation, $G_j$ there are
two right--moving, $U(1)_{r_j}$ and $U(1)_{r_{j+3}}$, symmetries.
For every right--moving $U(1)$ gauged symmetry, there is a corresponding
left--moving global $U(1)$ symmetry, $U(1)_{\ell_j}$ and
$U(1)_{\ell_{j+3}}$. Each sector $b_1$, $b_2$ and $b_3$
has two Ising model operators,
($\sigma_4$, $\sigma_5$), ($\sigma_2$, $\sigma_6$) and
($\sigma_1$, $\sigma_3$),
respectively, obtained by pairing a left--handed real fermion with
a right--handed real fermion.

The Neveu--Schwarz (NS) sector corresponds to the untwisted sector of the
orbifold model and
produces in addition to the gravity and gauge multiplets three pairs of
electroweak
scalar doublets $\{h_1, h_2, h_3, {\bar h}_1, {\bar h}_2, {\bar h}_3\}$,
three pairs of $SO(10)$ singlets with $U(1)$ charges,
$\{\Phi_{12},\Phi_{23},\Phi_{13},{\bar\Phi}_{12},
{\bar\Phi}_{23}, {\bar\Phi}_{13}\}$, and three scalars that are singlets
of the entire four dimensional gauge group, $\xi_1,\xi_2,\xi_3$.

The sector ${S+b_1+b_2+\alpha+\beta}$ ($\alpha\beta$ sector) also
produces states that transform only under the observable gauge group.
In addition to one pair of
electroweak doublets, $h_{45},{\bar h}_{45}$, and one pair of color
triplets, there are seven pairs of $SO(10)$
singlets with horizontal $U(1)$ charges, $\{\Phi_{45},{\bar\Phi}_{45},
\Phi_{1,2,3}^\pm,{\bar\Phi}_{1,2,3}^\pm\}$.

In addition to the states from these sectors, which transform solely under
the observable gauge group, the sectors $b_j+2\gamma$ produce states which are
$SO(10)$ singlets and transform as the $16$ vector representation of
the hidden $SO(16)$, decomposed under $SU(5)\times SU(3)\times U(1)^2$,
$\{T_{1,2,3},{\bar T}_{1,2,3},V_{1,2,3},{\bar V}_{1,2,3}\}$. The $T_i$
(${\bar T}_i$) are $5$ (${\bar 5}$) and the $V_i$ (${\bar V}_i$) are
$3$ (${\bar 3}$) of the hidden $SU(5)$ and $SU(3)$ gauge groups
respectively (in order not to cause any confusion between the hidden sector
states $T_i$ and the moduli we will call the moduli $\hat T_i$ throughout the
paper). These states arise due to the $Z_2\times Z_2$ twist
of $SO(12)\times SO(16)\times SO(16)$ rather than of
$SO(12)\times E_8\times E_8$, in  which they are replaced by $10+1$ of
$SO(10)$ and thus complete the 16 of $SO(10)$ to 27 of $E_8$. These
states are charged under the horizontal $U(1)_{r_j}$
symmetries and play an important role in generating quark flavor mixing.

The vectors with some combination of $(b_1,b_2,b_3,\alpha,\beta)\pm\gamma+(I)$
produce additional states in vector--like representations. Most of those
are Standard Model singlets but carry nonvanishing $U(1)_{Z^\prime}$
charge, where $U(1)_{Z^\prime}$ is the $U(1)$ inside
$SO(10)$ that is orthogonal to the electroweak hypercharge.
The sectors $b_{1,2}+b_3+\alpha\pm\gamma+(I)$ also produce
a pair of $SU(3)_C$ triplets and a pair of electroweak doublets.

The cubic level superpotential and higher order nonrenormalizable terms
in the superpotential are obtained by calculating correlators between
vertex operators,
$A_N\sim\langle V_1^fV_2^fV_3^b\cdot\cdot\cdot V_N^b\rangle,$
where $V_i^f$ $(V_i^b)$ are the fermionic (scalar)
components of the vertex operators.
The nonvanishing terms must be invariant under all the symmetries of the
string models and must satisfy all the string selection rules [\KLN].
To obtain the correct ghost charge, $(N_b-1)$ of the bosonic vertex
operators have to be picture changed from the $-1$ ghost picture to the
$0$ ghost picture. The invariance under the six global left--moving $U(1)$
symmetries and the Ising model correlators must be checked after all
picture changing operations have been performed. The invariant terms
are extracted by using a simple FORTRAN code.

The cubic level superpotential is given by [\EU],
$$\eqalignno{W&=\{(
{u_{L_1}^c}Q_1{\bar h}_1+{N_{L_1}^c}L_1{\bar h}_1+
{u_{L_2}^c}Q_2{\bar h}_2+{N_{L_2}^c}L_2{\bar h}_2+
{u_{L_3}^c}Q_3{\bar h}_3+{N_{L_3}^c}L_3{\bar h}_3)\cr
&\qquad
+{{h_1}{\bar h}_2{\bar\Phi}_{12}}
+{h_1}{\bar h}_3{\bar\Phi}_{13}
+{h_2}{\bar h}_3{\bar\Phi}_{23}
+{\bar h}_1{h_2}{\Phi_{12}}
+{\bar h}_1{h_3}{\Phi_{13}}
+{\bar h}_2{h_3}{\Phi_{23}}\cr
&\qquad
+{\Phi}_{12}(\Phi_{23}{\bar\Phi}_{13}+\Phi_1^-\Phi_1^+
+\Phi_2^-\Phi_2^+
+\Phi_3^-\Phi_3^+)\cr
&\qquad
+{\bar\Phi}_{12}({\bar\Phi}_{23}{\Phi}_{13}
+{\bar\Phi}_1^+{\bar\Phi}_1^-
+{\bar\Phi}_2^+{\bar\Phi}_2^-
+{\bar\Phi}_3^+{\bar\Phi}_3^-)\cr
&\qquad
+{1\over2}\xi_3(\Phi_{45}{\bar\Phi}_{45}
+\Phi_1^+{\bar\Phi}_1^++
\Phi_1^-{\bar\Phi}_1^-+\Phi_2^+{\bar\Phi}_2^++\Phi_2^-{\bar\Phi}_2^-
+\Phi_3^+{\bar\Phi}_3^++\Phi_3^-{\bar\Phi}_3^-\cr
&\qquad
+h_{45}{\bar h}_{45}+D_{45}{\bar D}_{45})
+h_3{\bar h}_{45}\Phi_{45}+{\bar h}_3h_{45}{\bar\Phi}_{45}\}\cr
&\qquad
+\{{1\over2}[\xi_1(H_{19}H_{20}+
H_{21}H_{22}+H_{23}H_{24}+H_{25}H_{26})\cr
&\qquad
+\xi_2(H_{13}H_{14}+H_{15}H_{16}+H_{17}H_{18})]
+{\bar\Phi}_{23}H_{24}H_{25}
+{\Phi}_{23}H_{23}H_{26}\cr
&\qquad
+h_2H_{16}H_{17}
+{\bar h}_2H_{15}H_{18}
+{e_{L_1}^c}{H_{10}}{H_{27}}
+{e_{L_2}^c}{H_8}{H_{29}}
+{H_{27}}({V_1}{H_9}\cr
&\qquad
+{V_2}{H_{11}})+{V_6}{H_5}{H_{29}}
+{{\bar\Phi}_{45}}{H_{17}}{H_{24}}
+{D_{45}}{H_{18}}{H_{21}}
+{h_{45}}{H_{16}}{H_{25}}\}\quad&(1)\cr}$$
where a common normalization constant ${\sqrt 2}g$ is assumed.
{}From Eq. (1) it is seen that only $+{2\over3}$ charged quarks
obtain a cubic level mass term. This result arises due to the assignment
of boundary conditions in the vector $\gamma$ [\SLM].
Mass terms for $-{1\over3}$ and for
charged leptons must be obtained from nonrenormalizable terms.

An important property of the superstring standard--like models is the
absence of gauge and gravitational anomalies apart from a single ``anomalous
$U(1)$" symmetry. This ``anomalous''
$U(1)_A$ generates a Fayet--Iliopoulos term
that breaks supersymmetry at the Planck scale [\DSW]. Supersymmetry is
restored and $U(1)_A$ is broken by giving
VEVs to a set of Standard Model singlets in the
massless string spectrum along the flat F and D directions [\ADS].
The $SO(10)$ singlet fields in the nonrenormalizable terms
obtain nonvanishing VEVs by the application of the DSW mechanism. Thus the
order $N$ nonrenormalizable terms, of the form $cffh(\Phi/M)^{N-3}$,
become effective trilinear terms, where $f,h,\Phi$ denote fermions, scalar
doublets and scalar $SO(10)$ singlets, respectively, and
$M=M_{Pl}/(2\sqrt{8\pi})\approx10^{18}GeV$.
The effective Yukawa couplings are given by
$\lambda=c(\langle \Phi \rangle/M)^{N-3}$, where the calculable coefficients
$c$ are of order one [\KLN], and are suppressed relative to the Yukawa
couplings that are obtained at lower orders (since $\l \Phi \r \sim M/10$
generically). In this manner quark mass terms,
as well as quark mixing terms, can be obtained. Realistic quark masses and
mixing can be obtained for a suitable choice of scalar VEVs.
Requiring quark mixing angles of the correct order of magnitude
necessitates that we give nonvanishing VEVs
to some states from the sectors
$b_j+2\gamma$ in the application of the DSW mechanism.

The form of the effective supergravity theory in the free fermionic
model was analyzed in Ref. [\AEFNT,\FGKP].
The K\"ahler potential takes the form
$$K=-\ln{(S+S^\dagger)}-\sum_{a=1}^3{\ln{{\eta_a}^0}}+\sum_\alpha{2\ln
{(1-{\phi_\alpha}{{\phi_\alpha}^\dagger})}}~,\eqno(2a)$$
where
$$\eqalignno{
&{\eta_a}^0={1\over\sqrt2}\left\vert1+\vert{\hat\eta_a}^1\vert^2-
\sum_{i_a=2}^{n_a}\vert{\hat\eta_a}^{i_a}\vert^2\right\vert^{1\over
2}~,&(2b)\cr
&{\hat\eta}_a^{1^2}=-1+\sum_{i_a=2}^{n_a}({\hat\eta_a}^{i_a})^2,&(2c)}$$
where $S$ and $\phi_i$ are the dilaton and chiral matter fields and
$(\eta_a^0,\eta_a^1,\eta_a^{i_a})$ correspond to the
physical and unphysical components of the moduli fields [\AEFNT,\FGKP].
The value of $e^K$ is determined by the expectation values of $S$,
of the gauge non--singlet ${\hat\eta_a}^{i_a}$ and of the matter fields.
In the free fermionic models, the fields that
correspond to the moduli fields in orbifold models, appear
in D--terms and possibly have nontrivial superpotential couplings.
Their VEVs are subject to the F and D--flatness constraints.

To study the problem of supersymmetry breaking, we have to relate the
effective field theory that is obtained in the free fermionic
models to the one that is used in supergravity.
In most analyses of supersymmetry
breaking in superstring inspired field theories, one assumes the
K\"ahler potential at tree level to take the form:
$$K=-\ln{(S+{S^\dagger})}-3\ln{(\hat T+{\hat T^\dagger}-\sum_i {{\vert{\phi_i}
\vert}^2})},\eqno (3)$$
where $S$ and $\hat T$ are the dilaton and overall
modulus field and the $\phi_i$ are the
gauge nonsinglet chiral superfields.
The effective supergravity action is invariant under discrete target space
duality transformations [\GPR],
$$\hat T_j \to {{a_j {\hat T_j}-ib_j}\over {ic_j{\hat T_j}+d_j}}, \eqno(4)$$
where $j=1,2,3$ and $a_j,b_j,c_j,d_j$ are integers which satisfy
$a_jd_j-b_jc_j=1$.
This symmetry has been used extensively to constrain
the superstring inspired effective field theory and was
also discussed in S-matrix approach to the effective string field
theory [\DKL]. In the free fermionic models,
if we assume that the fields that correspond to the moduli are correctly
identified, then we may assume a K\"ahler potential of the form of Eq. (3).
The modular group of the free fermionic models is $PSL(2,Z)^3$ and
the associated moduli fields are denoted by $\hat T_j$ $(j=1,2,3)$. The modular
weight of each matter field is related to the charges under the
left--moving $U(1)_{\ell_{1,2,3}}$ symmetries, 
where $U(1)^\prime=U(1)_1+U(1)_2+U(1)_3$ is the $U(1)$ current in the
left--moving
$N=2$ world--sheet supersymmetry. The matter fields then transform as
$$\phi_i\rightarrow \phi_i\prod_j(ic_j{\hat T_j}+d_j)^{-w^i_j},\eqno(5)$$
under modular
transformations, where ${-w^i_j}$ are the charges of the scalar
component of the $\phi_i$ superfields under the left--moving $U(1)_{\ell_j}$
$(j=1,2,3)$ symmetries.  Under target space duality transformation the
superpotential has overall weight $-3$ and the dilaton has weight zero.
Modular invariance is manifest at the cubic level of the superpotential.
The existence of a nonvanishing term at some
order $N$ implies the existence of an infinite number of nonrenormalizable
terms. These terms are obtained by tagging,
to the order $N$ term, powers of scalar fields in the massless
spectrum that correspond to the moduli fields [\KLNII].
Tagging moduli to the order $N$ term corresponds to multiplying the
order $N$ term by an appropriate power of the Dedekind $\eta$ function,
with modular weight $1/2$, which renders the theory with the
nonrenormalizable terms modular invariant.
However, as the corrections from the infinite tower of nonrenormalizable
terms are suppressed by powers of $({{\l \phi \r}/M})^n$, for our purposes
we will neglect the modular invariance
of the effective theory, asserting that the full theory will only induce
small corrections to our results.
Thus, we use the following K\"ahler function for our calculation:
$$K=-\ln{(S+{S^\dagger})}-{\sum_{a=1}^3}{\ln{{\eta_a}^0}}+f(\hat T,{\hat
T^\dagger})\phi_i \phi_i^{\dagger},\eqno(6)$$
where $f(\hat T,{\hat T^\dagger})$ is a function that may depend on the moduli.

\bigskip
\centerline{\bf 3. Cubic level F and D-constraints }

We now proceed to analyze the F and D-flatness constraints in the cubic level
superpotential. The set of F and D constraints is given by the following
equations:
$$\eqalignno{{D_A}&={\sum _{k}}{Q_k^A}{\vert\chi_k\vert}^2=
{-g^2e^{\phi_D}\over{192\pi^2}}{1\over{2 \alpha^{\prime}}}{Tr(Q_A)}&(7a)\cr
D_j'&=\sum_{k}{Q'}_k^j\vert\chi_k\vert^2=0{\hskip .3cm}
 j=1\cdots5&(7b)\cr
D_j&=\sum_{k}Q_k^j\vert\chi_k\vert^2=0{\hskip .3cm} j=C,L,7,8&(7c)\cr
W&={{\partial W}\over{\partial{\eta_i}}}=0&(7d)\cr}$$
where $\chi_k$ are the fields that get a VEV and $Q_k^j$ is their charge
under the $U(1)_j$ symmetry. The set
$\{{\eta}_i\}$ is the set of fields with vanishing VEV.
$\alpha^\prime$ is the string
tension and $1/(\sqrt{2\alpha^\prime})=gM_{Pl}/(2\sqrt{8\pi})=gM
\sim 10^{18}~GeV$. In the model of Ref. [\EU], $Tr(Q_A)=180$.

The observable sector F flatness conditions derived from the cubic
superpotential are:
$$\eqalignno{
{{\bar\Phi}_{13}}{\Phi_{12}}&+{H_{23}}{H_{26}}=
{{\Phi}_{13}}{\bar\Phi_{12}}
+{H_{24}}{H_{25}}=0 &(8a)\cr
{{\Phi}_{23}}{\Phi_{12}}&={{\bar\Phi}_{23}}
{\bar\Phi_{12}}=0&(8b)\cr
{{\bar\Phi}_{23}}{\Phi_{13}}&+{\bar\Phi}^+_i{\bar\Phi}^-_i=0&(8c)\cr
{\bar\Phi}_i^+{\bar\Phi}_{12}&+\Phi_i^-\xi_3=0&(8d)\cr
{\bar\Phi}_i^-{\bar\Phi}_{12}&+\Phi_i^+\xi_3=0&(8e)\cr
\Phi_{45}{\bar\Phi}_{45}&+\Phi_i^+{\bar\Phi}_i^+
+\Phi_i^-{\bar\Phi}_i^-=0&(8f)\cr
\Phi_{45}\xi_3&+H_{17}H_{24}=0&(8g)\cr
\bar\Phi_{45}\xi_3&=0.&(8h)\cr}$$
For equations $(8c)-(8e)$ the barred equations have to be taken
as well.
{}From the $\l D_A\r=0$ constraint we can show  that $\Phi_{45}$ must
get a nonvanishing VEV. We can always modify the set of $D$ constraints
in a way that only the coefficient of $\l\Phi_{45}\r$
is different from zero and
consequently $\Phi_{45}$ must get a VEV to cancel the ``anomalous'' $U(1)$
D--term equation. From Eq. (8g) it follows that $\l \xi_3 \r =0$ because
$\l H_{17} \r=0$, as we will see below. In general,
some additional $\Phi_i^\pm$ and ${\bar\Phi}_i^\pm$ obtain VEVs from the set
of D--term constraints. Therefore, from Eqs. (8d-8e) it follows that
$\l \Phi_{12} \r=0$ and $\l {\bar \Phi}_{12} \r=0$.
We conclude that the cubic level F--flatness constraints in the observable
sector impose
$$\l {\Phi_{12}},{\bar \Phi_{12}},{\xi_3}\r=0.\eqno(9)$$

In the hidden sector, on the other hand, we get the following F constraints
from the cubic superpotential:
$$\eqalignno{&H_{19}H_{20}+H_{23}H_{24}+H_{25}H_{26}=0 &(10a)\cr
            &H_{13}H_{14}+H_{17}H_{18}=0 &(10b)\cr
            &{1\over 2}\xi_1H_{24}+\Phi_{23}H_{26}=0 &(10c)\cr
            &{1\over 2}\xi_1H_{23}+\bar \Phi_{23}H_{25}=0 &(10d)\cr
            &{1\over 2}\xi_1H_{26}+ \bar \Phi_{23}H_{24}=0 &(10e)\cr
            &{1\over 2}\xi_1H_{25}+\Phi_{23}H_{23}=0 &(10f)\cr
            &{1\over 2}\xi_2H_{13}={1\over 2}\xi_2H_{14}={1\over 2}\xi_2H_{17}=
{1\over 2}\xi_2H_{18}=0.  &(10g)} $$
The requirement of realistic (or nonzero) $b$, $s$, $\mu$, $\tau$ masses
imposes that $\xi_1$ and $\xi_2$ must get VEVs [10,12].
Below we will show that $\l\xi_1,\xi_2\r\ne0$ also insures the stability
of the supersymmetric vacuum.
Then, from Eq. (10g) we obtain
$$\l H_{13} \r=\l H_{14} \r=\l H_{17} \r=\l H_{18} \r=0, \eqno(11)$$
in order to preserve supersymmetry at $M_{Pl}$. $H_{19}$ and $H_{20}$ are
$5$ and $\bar5$ of $SU(5)$. In the scenario that we study we assume that
the hidden $SU(5)$ group is unbroken at $M_{Pl}$,
which imposes $\l H_{19} H_{20} \r=0$.
For the rest, $H_{23},H_{24},H_{25},H_{26}$, from Eqs. (10c--f) we get the
following constraints:

a) When $\l \Phi_{23} \r \not =0$ and $\l \bar \Phi_{23} \r \not =0$, either
$\l H_{23} \r=\l H_{25} \r=0 $ or $\l H_{24} \r= \l H_{26} \r=0 $.

b) When one or both of $\Phi_{23},\bar \Phi_{23}$ have vanishing VEVs,
$\l H_{23} \r=\l H_{25} \r =\l H_{24} \r= \l H_{26} \r=0 $.

We stress that these supersymmetry constraints on the hidden sector VEVs are
obtained by requiring a realistic heavy quark and lepton spectrum
(i.e. $\l \xi_{1,2}
\r \not =0$) and a nonsingular hidden matter mass matrix. Otherwise,
Eqs. (10a--g) do not lead to useful supersymmetry constraints on the hidden
sector VEVs.

Given the assumption that VEVs that break $U(1)_{Z'}$, i.e. all the $H_i$,
are suppressed
the set of fields that may obtain VEVs is divided into two classes:

\parindent=-15pt

1. Singlets from the NS sector and the sector $S+{b_1}+{b_2}+{\alpha}+{\beta}$.
These states transform solely under the observable gauge group and are
$SO(10)\times SO(16)$
singlets.

2. States from the sectors ${b_j}+2\gamma$, $j=1,2,3$.
These states are obtained from the twisted sectors of the orbifold model.
They are $SO(10)$ singlets and transform
under the hidden gauge group as the 16 representation
of $SO(16)$ decomposed under $SU(5)\times {SU(3)}\times {U(1)^2}$.

\parindent=15pt

If we restrict the allowed VEVs only to states from the first class,
then the cubic
level F--flat solution is preserved to all orders of nonrenormalizable terms.
This result follows from the charges of the states from these sectors under the
left--moving global $U(1)_{\ell_{1,2,3}}$ symmetries.
The order $N$ terms that have to be investigated are of the form,
$$\langle(\alpha\beta)^j(NS)^{N-j}\rangle{\hskip2cm}
(j=4,\cdots,N),\eqno(12)$$
where (NS) denotes fields from the Neveu--Schwarz sector
and $(\alpha\beta)$ denotes fields
that belong to the sector ${b_1}+{b_2}+\alpha+\beta$.
Without loss of generality, we can choose two of the $(\alpha\beta)$ fields
to be the two space--time fermions in these correlators.
The $U(1)_{\ell_{1,2,3}}$ charges for the $(\alpha\beta)$ fields are
$(0, 0, {1\over 2})$ for the
fermions and $(-{1\over 2}, -{1\over 2}, 0)$ for the bosons.
All NS fields in Eq. (12) are bosonic fields with charges
$U(1)_{\ell_j}=0$ or $-1$.
Of the NS singlets, only $\Phi_{12}$, ${\bar\Phi_{12}}$
and $\xi_3$ carry $U(1)_{\ell_3}$ charges.
We can always choose a basis in which the $U(1)_{\ell_3}$ charge of these
fields is picture changed to zero.
The picture changing operation on the $(\alpha\beta)$ scalars can only change
them to $(\pm{1\over 2}, \pm {1\over 2}, 0)$.
Therefore, none of the terms of the form of Eq. (12) are invariant under
$U(1)_{\ell_3}$. The conclusion is that all these terms vanish identically
to all orders.
Thus, if we allow only VEVs for the states from the NS sector and the
sector ${b_1}+{b_2}+\alpha+\beta$, there are no new F terms
even if we include nonrenormalizable terms to all
orders. Consequently, in this case, the cubic level F and D--flat solution
is valid to all orders $N$ and supersymmetry is unbroken.

However, to obtain a Cabibbo angle of the correct order of magnitude it is
necessary to give VEVs to some states from the sector ${b_j}+2\gamma$.
This result follows from the following considerations.
The Higgs doublets $h_3$ and $\bar h_3$ obtain a Planck scale mass
due to the cubic level F and D--flatness constraints [\NRT,\FM].
As a result, and because the horizontal charges forbid
the states from the sector $b_3$ to couple
directly to the remaining Higgs doublets, the states from the
sector $b_3$ are identified with the lightest generation, while the
sectors $b_1$ and $b_2$ are identified with the two heavier
generations. However, due to the $U(1)_{\ell_{j+3}}$ horizontal
symmetries, mixing terms between the sectors
$b_1$, $b_2$ and $b_3$ are only obtained by exchanging states from the
sector ${b_j}+2\gamma$ [\FM].
Consequently, to obtain a Cabibbo angle that is not too small
necessitates that we give VEVs of order $M/10$ to some states from the sector
${b_j}+2\gamma$ [\CKM].
Thus, when analyzing the effective supergravity theory that is obtained
from these models, we have to include these hidden sector VEVs as well.
In the next section, we show that the same VEVs must be imposed to obtain
a nonsingular mass matrix for the hidden $SU(5)$ matter states which is
essential for a stable supersymmetric vacuum [\SQCD].
Thus, the same VEVs that generate the Cabibbo mixing also guarantee
that the supersymmetric vacuum is well defined.

We therefore have to find F and D flat solutions that
contain nonvanishing VEVs for the states from the sectors $b_j+2\gamma$.
In the following,
to illustrate our scenario,
we consider a specific F and D flat solution,
which corresponds to a specific choice of string vacuum at
$M_{Pl}$. We do not know the string dynamics that select the string
vacuum but simply consider one.
An explicit solution that satisfies all the cubic level F and D
flatness constraints is
given by the following set of nonvanishing VEVs,
$$\{{\bar V}_2,{V}_3,\Phi_{45},{\bar\Phi}_{13},{\bar\Phi}_1^-,
\Phi_2^+,{\bar\Phi}_3^-,\xi_1,\xi_2\},\eqno(13)$$
with
$$\eqalignno{2\vert{\l{\bar V}_2\r}\vert^2
=2\vert\l{V}_3\r\vert^2={1\over4}
\vert\l\Phi_{45}\r\vert^2=\vert\l{\bar\Phi}_{13}\r\vert^2&=
{{g^2}\over{16\pi^2}}{1\over{2\alpha^\prime}},&(14a)\cr
2\vert\l{\bar\Phi}_1^-\r\vert^2=
2\vert\l{\Phi_2^+}\r\vert^2=\vert\l{{\bar\Phi}_3^-}\r\vert^2&=
{{g^2}\over{16\pi^2}}{1\over{2\alpha^\prime}},&(14b)\cr}$$
The VEVs of  $\xi_1$ and $\xi_2$ are not fixed
by the F and D constraints. In general
$\l\xi_1,\xi_2\r={\cal O}(g^2M/4\pi)$ must be imposed to
obtain realistic quark and lepton masses [\NRT,\FM]. The fields $\{\Phi_{12},
{\bar\Phi}_{12},\Phi_{13},{\bar\Phi}_{13},\Phi_{23},{\bar\Phi}_{23}\}$
are related to the untwisted moduli fields of the $Z_2\times Z_2$ orbifold.
Their VEVs in the fermionic model are determined by the F and D--flatness
requirements. From Eq. (14a) and with
$\vert\l {\bar\Phi}_{23} \r\vert^2=\vert\l {\hat\eta}_1 \r\vert^2$,
$\vert \l{\bar\Phi}_{13} \r\vert^2=\vert \l{\hat\eta}_2 \r\vert^2$, and
$\vert \l{\bar\Phi}_{12} \r\vert^2=\vert \l{\hat\eta}_3 \r\vert^2$,
it is seen that we have $\vert \l{\hat\eta}_j \r\vert^2=\{
0,g^2/(16\pi^2),0\}$, in units of $M$.
Inserting these values into Eq. (2) for the
K\"ahler function, we observe that $\l \eta_j^0 \r \approx\sqrt2,~ j=1,2,3$.
Hence, the analogue of the conventional moduli fields are determined to be of
order one.

The hidden sector of the superstring model contains two non--Abelian gauge
groups, $SU(3)$ and $SU(5)$. The $SU(3)$ group is broken by our choice of
VEVs and does not play a role in our analysis of supersymmetry breaking.
The massless spectrum contains four pairs of $5+\bar5$ of the hidden $SU(5)$
gauge group. Of those $H_{19}$ and $H_{20}$ obtain Planck scale masses due
to the cubic level term $H_{19}H_{20}\xi_2$. Thus we obtain below the Planck
scale a $SU(5)$ gauge group with three pairs of $5+\bar5$.

\bigskip
\centerline{\bf 4. Supersymmetry breaking from nonrenormalizable terms}

In the previous section we showed that the solution to the cubic level
F and D constraints divide into two classes:
those that include states from the
sectors $b_j+2\gamma$ and those that do not.
For the second class of solutions,
there are no corrections from nonrenormalizable terms to the cubic level
F and D--flat solution.
However, the second class of solutions cannot produce realistic quark
mixing. Therefore, the
realistic solution can only be of the first class.
Furthermore, the unbroken non--Abelian hidden sector
gauge groups (which in our case is hidden $SU(5)$) condense at some scale
$\Lambda_H$. Then hidden matter condensates
foenormalizable terms
in the superpotential and taking into
account the nonperturbative effects of the hidden
$SU(5)$ gauge group. We find that
contrary to the second class solutions, in this case
the nonrenormalizable terms modify the cubic level constraints.
We investigate this
modification and argue that when nonrenormalizable terms with hidden sector
condensates are taken into account
the set of F constraints cannot be satisfied
simultaneously. As a result nonvanishing
F--terms are generated and supersymmetry
is broken.

We search for allowed nonrenormalizable terms that contain $SO(10)$
singlet fields from the NS, $\alpha\beta$  and $b_j+2\gamma$ sectors.
At order $N=5$ we find,
$$\eqalignno{
  &V_2{\bar V_2}\Phi_{45}\Phi_2^-\xi_1,{\hskip 2cm}
   V_1{\bar V_1}\Phi_{45}{\bar\Phi}_1^+\xi_2,&(15a,b)\cr
  &T_2{\bar T_2}\Phi_{45}\Phi_2^+\xi_1,{\hskip 2cm}
   T_1{\bar T_1}\Phi_{45}{\bar\Phi}_1^-\xi_2,&(15c,d)\cr}$$
at order N=7,
$$\eqalignno{
  &T_2{\bar T_3}V_3{\bar V_2}\Phi_{45}\Phi_{45}{\bar\Phi}_{13},&(16a)\cr
  &T_1{\bar T_2}V_1{\bar V_2}\Phi_{45}\Phi_{45}\xi_1,&(16b)\cr
  &T_2{\bar T_1}V_1{\bar V_2}\Phi_{45}\Phi_{45}\xi_2,&(16c)\cr}$$
$$\eqalignno{
  &V_2{\bar V_2}\Phi_{45}\Phi_2^-\xi_1
[({{\partial W_3}\over{\partial\xi}_3})+\xi_i\xi_i+
\Phi_{13}{\bar\Phi}_{13}+\Phi_{23}{\bar\Phi}_{23})],&(16d)\cr
  &V_1{\bar V_1}\Phi_{45}{\bar\Phi}_1^+\xi_2
[({{\partial W_3}\over{\partial\xi}_3})+\xi_i\xi_i+
\Phi_{13}{\bar\Phi}_{13}+\Phi_{23}{\bar\Phi}_{23})],&(16e)\cr
&V_2{\bar V_2}\Phi_{45}{\bar\Phi}_2^+\xi_1
({{\partial W_3}\over{\partial\Phi}_{12}}),&(16f)\cr
&V_1{\bar V_1}\Phi_{45}{\Phi}_1^-\xi_2
({{\partial W_3}\over{\partial{\bar\Phi}_{12}}}),&(16g)\cr}$$
and at order N=8,
$$\eqalignno{&{\bar T_2} T_3 V_3{\bar V_2}\Phi_{45}\Phi_{45}{\bar\Phi}_{13}
\xi_1, &(17a)\cr
& T_1 {\bar T_3} V_1{\bar V_3}\Phi_{45}\Phi_{45}{\bar\Phi}_{23}
\xi_2, &(17b)\cr
&T_2 {\bar T_3} V_2{\bar V_3}\Phi_{45}\Phi_{45}{\bar\Phi}_{13}
\xi_1, &(17c)\cr
&T_3 {\bar T_1} V_3{\bar V_1}\Phi_{45}\Phi_{45}{\bar\Phi}_{23}
\xi_2. &(17d)}$$

First, we examine the terms that include in addition to NS and $\alpha\beta$
states also $V_{1,2,3}$ and ${\bar V}_{1,2,3}$. To produce realistic
quark mixing some $V_j$ and ${\bar V}_j$ must
obtain VEVs in the cancelation of the ``anomalous'' $U(1)$ D--term equation.
Eqs. (15a,b) indicate that giving VEVs to
both $V_j$ and ${\bar V}_j$ from a sector
$b_j+2\gamma$ may result in too large a supersymmetry
breaking scale in the observable sector.
Therefore, we require that the order $N=5$
terms, Eqs. (15a,b), vanish as well as all of their derivatives.
This imposes additional constraints on the allowed
VEVs, namely, $\l\Phi_2^-\r=0$ and $\l{\bar\Phi}_1^+\r=0$ and either $V_j$ or
${\bar V}_j$ from each sector $b_j+2\gamma$ can get a VEV.
Then, the $N=5$ and $N=7$ order terms, of the form $V_j \bar V_j \phi^n$, as
well as all of their derivatives, vanish due to our choice of flat
directions and the cubic level constraints
$({{\partial W_3}/{\partial{\phi_i}}})=0$.
We then find that there are no additional nontrivial constraints from
nonrenormalizable terms up to order $N=11$.

Next we examine the terms that contain $5$ and $\bar 5$ of the hidden $SU(5)$
gauge group, {\it i.e.} $T_{1,2,3}$ and ${\bar T}_{1,2,3}$.
The scale at which
the hidden $SU(5)$ gauge group becomes strongly interacting is given by
$${\Lambda_5}=M\exp({{2\pi}\over{b}}{{(1-\alpha_0)}\over
\alpha_0}),\eqno(18)$$
where $b={1\over2}n_5-15$. For $n_5=6$ and $\alpha_0=(1/25-1/20)$,
$\Lambda_5\sim(10^{12}-10^{14})GeV$.
The value of $\alpha_0$ assumes that the dilaton
potential has a minimum with $\l S\r\sim 1/2$. The $T$'s and
$\bar T$'s will form hidden sector condensates when the hidden $SU(5)$
gauge group becomes strongly interacting. To evaluate the gaugino and matter
condensates we use the well known expressions for supersymmetric $SU(N)$
with matter in $N+{\bar N}$ representations [\SQCD],
$$\eqalignno{
{1\over{32\pi^2}}\vev{\lambda\lambda}
&=\Lambda^3\left(det{{\cal M}\over\Lambda}\right)^{1/N},&(19a)\cr
\Pi_{ij}=\vev{{\bar T_i}T_j}&={1\over{32\pi^2}}\vev
{\lambda\lambda}{{\cal M}_{ij}}^{-1},&(19b)\cr}$$
where $\l\lambda\lambda\r$, $\cal M$ and
$\Lambda$ are the hidden gaugino condensate, the hidden matter mass matrix
and the $SU(5)$ condensation scale, respectively.
Modular invariant generalization of Eqs. (19a,b) for the string case were
derived in Ref. [\LT].
The nonrenormalizable terms can be put in modular invariant form
by following the procedure outlined in Ref. [\KLNI].
Approximating the Dedekind $\eta$ function by $\eta({\hat T})\approx
e^{-\pi {\hat T}/12}(1-e^{-2\pi {\hat T}})$
we verified that the calculation using the modular invariant expression
from Ref. [\LT] (with $\l {\hat T}\r\approx M$) differ from the results using
Eq. (19), by at most an order of magnitude. Therefore, for the
purpose of our qualitative observations the use of Eqs. (19a,b) is
adequate.

In our case the matrix $\cal M$ is given by,
$${\cal M}=\left(\matrix{ 0   & C_1   & 0   \cr
			  B_1 & A_2   & C_2 \cr
			  0   & C_3   & A_1   \cr  }\right)~,\eqno(20)$$
where $A,B,C$ arise from terms at orders $N=5,8,7$ respectively and are
given by
$$\eqalignno{A_1&= {{\l\Phi_{45}{\bar\Phi}_1^-\xi_2\r}\over{M^2}},
\qquad \qquad A_2= {{\l\Phi_{45}\Phi_2^+\xi_1\r}\over{M^2}}
		   						,&(21a,b)\cr
B_1&= {{\l V_3{\bar V_2}\Phi_{45}\Phi_{45}{\bar\Phi}_{13}\xi_1\r}\over
{M^5}},&(21c)\cr
C_1&= {{\l V_3{\bar V_2}\Phi_{45}\Phi_{45}{\bar\Phi}_{13}\r}\over{M^4}},~~
C_2 = {{\l V_1{\bar V_2}\Phi_{45}\Phi_{45}\xi_1\r}\over{M^4}},&(21d,e)\cr
C_3&= {{\l V_1{\bar V_2}\Phi_{45}\Phi_{45}\xi_2\r}\over{M^4}}~~.
&(21f)}$$

For the solution given by Eq. (14) $V_1=0\Rightarrow C_2=C_3=0$.
Taking generically $\l\phi\r\sim{gM}/4\pi\sim M/10$ we obtain
$A_i \sim 10^{15}~GeV$, $B_i \sim 10^{12}~GeV$, and $C_i\sim 10^{13}~GeV$.
The matrix ${\cal M}^{-1}$ is given by,
$${\cal M}^{-1}=
\left(\matrix{
 -{{{A_1A_2} - {C_1C_2}}\over {A_1\,B_1\,C_1}}
& {1\over B_1} & -{C_2\over {A_1\,B_1}}\cr
  {1\over C_1} & 0 & 0 \cr
 -{C_3\over {A_1C_1}} & 0 & {1\over A_1}\cr}\right)~.\eqno(22)$$
The zeros in ${\cal M}^{-1}$ indicate that of the above
nonrenormalizable terms only the ones with
$T_1{\bar T}_1$ and $T_2{\bar T}_3$ and
${\bar T}_2T_3$ remain and the rest vanish.
For the specific solution that we
consider $\Pi_{13}$ and $\Pi_{31}$ vanish as well.
This does not affect our results
because these condensates do not appear in
the nonrenormalizable terms  at least up to order
$N=8$. To obtain a stable vacuum the hidden matter
mass matrix must be nonsingular, {\it i.e.} the
determinant of $\cal M$ must not vanish [25]. To obtain a nonsingular
determinant we must impose some nonvanishing VEVs. From Eqs. (20) and (21)
we find that in particular we must impose that $\bar V_2$ and $V_3$,
as well as $\{\Phi_{45},\Phi_2^+,{\bar\Phi}_1^-,{\bar\Phi}_{13},\xi_{1,2}\}$,
get nonvanishing VEVs.
We observe that a stable supersymmetric vacuum requires hidden sector VEVs.
It is interesting to note that the same constraint that must be imposed
to obtain realistic quark mixing
has to be imposed to obtain a stable supersymmetric vacuum.
With the cubic level solution given by Eq. (14),
${\cal M}$ is nonsingular
and its determinant is equal to $\hbox{Det}{\cal M}=-A_1B_1C_1$.

We are now ready to estimate the scale of supersymmetry breaking. Neglecting
the effects of the electroweak symmetry breaking, we have
$$V_{eff}= e^G (G_i^j)^{-1} G_j G^i-3e^G ,\eqno(23)$$
where $G=K+ln \vert W \vert  ^2$ and $K$ and $W$ are
given by Eqs. (6) and (1, 12--14)
respectively. Here the subscript (superscript) $i$ denotes
$\partial/\partial \phi_i$ ($\partial/\partial \phi_i^{\dagger}$).
Then,
$$V_{eff}= (G^i_i)^{-1} \vert F_i \vert ^2- 3e^K \vert W \vert ^2,\eqno(24)$$
where
$$ F_i=e^{K/2}( W_i+ W K_i),\eqno(25)$$
In general, inclusion of the nonrenormalizable terms in the superpotential
and condensation in the hidden sector
modifies the effective cubic superpotential.
As a result, there is a new vacuum corresponding
to the new effective potential.
The old F and D flat solution given by Eq. (14) will no longer correspond to
a supersymmetric minimum. In fact, now the solution given by Eq. (14) is
neither supersymmetric nor a minimum of the new effective potential.
To demonstrate explicitly that supersymmetry is broken, we would have
to write down the effective superpotential for all the fields in the massless
spectrum, minimize the effective potential and show that in the vacuum
one of the F--terms, Eq. (25), is different from zero. In this case
supersymmetry is spontaneously broken in the vacuum
of $V_{eff}$. However, due to the large number of fields and the
complicated superpotential, this is an
intractable task to perform analytically.
A numerical analysis has to be employed and is deferred to future work.
Instead, we resort to the following argument to show that supersymmetry
is broken in the new vacuum. The nonrenormalizable terms modify the cubic
level superpotential. We first focus on the modification due to the
quintic order terms. From Eq. (15) and Eq. (22) we observe that at the
quintic order there is a single term that modifies the cubic
level superpotential, $W_5=c\Pi_{11}\Phi_{45}{\bar\Phi}_1^-\xi_2.$
The coefficient $c$ is expected to be of order one [\KLN].
This term modifies the cubic level superpotential and the equations
for ${\partial W}/{\partial{\Phi}_{45}}$,
${\partial W}/{\partial{\bar\Phi}_1^-}$,
${\partial W}/{\partial\xi_2}$ where now $W=W_3+W_5$.

We see that $\l W_5 \r \not=0$ for the explicit F and D flat direction given
by Eq. (14). (We remind that with Eq. (14) $\l W_3 \r=0$).
$\l W_5 \r$ is necessarily nonvanishing for the following reasons.
It is always possible to rewrite the D--term equations in a way that the
only nonvanishing coefficient in the equation for $D_A$ is the
coefficient of $\vert\l\Phi_{45}\r\vert^2$. Thus, $\l\Phi_{45}\r$
must be different from zero. In addition, if $\l{\bar\Phi}_{1}^-\r$
and $\l\xi_2\r$ vanish the hidden matter mass matrix is singular and
the vacuum exhibits a runaway
behavior [\SQCD]. If we insist that the determinant of the matter condensates
is nonvanishing so that the vacuum is well defined, we find that
$\l W_3+W_5 \r \not=0$.

The modified F equations for $W=W_3+W_5$ read,
$$\eqalignno{
{{\partial W}\over{\partial{\bar\Phi}_1^-}}&=\bar \Phi_1^+{\bar\Phi}_{12}+
{1\over2}\xi_3\Phi_1^-+c\Pi_{11}\Phi_{45}\xi_2+c\Phi_{45}{\bar\Phi}_1^-\xi_2
{\partial\Pi_{11} \over \partial{\bar \Phi_1^-}}=0~,&(26a)\cr
{{\partial W}\over{\partial{\Phi}_{45}}}&={1\over2}\xi_3{\bar\Phi}_{45}+
c\Pi_{11}{\bar\Phi}_1^-\xi_2+c
\Phi_{45}{\bar\Phi}_1^-\xi_2 {\partial \Pi_{11} \over
\partial{\Phi}_{45}}=0~,&(26b)\cr
{{\partial W}\over{\partial\xi_2}}&=H_{13}H_{14}+H_{17}H_{18}+
c\Pi_{11}{\bar\Phi}_1^-\Phi_{45}+
c\Phi_{45}{\bar\Phi}_1^-\xi_2 {\partial \Pi_{11} \over
{\partial\xi_2}}=0~.&(26c)\cr}$$

The first two terms in these equations arise from the cubic level
superpotential whereas the
last two give the corrections from $W_5$. The last term in each equation
arises due to the implicit field dependence of the matter condensate
$\Pi_{11}$. $\Pi_{11}$ depends on the fields in two ways as can be seen from
Eq. (19b): through the gaugino condensate $\l \lambda \lambda \r$ which depends
on $det {\cal M}$ and through ${\cal M}^{-1}_{ij}$ given by Eq. (22).
The other cubic level constraints given by Eqs. (8) and (10) remain intact.
An explicit calculation of the last term in each equation
shows that the corrections to the F equations due to $W_5$ are given by
$$\eqalignno{
{{\partial W_5}\over{\partial{\bar\Phi}_1^-}}&={1\over5} \Pi_{11}\Phi_{45}
\xi_2, &(27a) \cr
{{\partial W_5}\over{\partial \Phi_{45}}}&=\Pi_{11} \bar \Phi_1^- \xi_2,
&(27b) \cr
{{\partial W_5}\over{\partial{\xi_2}}}&={1\over5} \Pi_{11}
\Phi_{45} \bar \Phi_1^-. &(27c)}$$
We see that the last two terms in each equation are nonzero.
The other cubic level equation remain intact. The cubic level
constraint Eq. (8g) and Eq. (8d,e)
still require $\l\Phi_{12}\r=\l{\bar\Phi}_{12}\r=
\l\xi_3\r=0$. But then the equations obtained from the quintic
order modification cannot be satisfied, unless
$\l\Phi_{45}\r=\l{\bar\Phi}_{1}^-\r=\l\xi_2\r=0$.
However, it is always possible to rewrite the D--term
equations in a way that the
only nonvanishing coefficient in the equation for $D_A$ is the
coefficient of $\vert\l\Phi_{45}\r\vert^2$. Thus, $\l\Phi_{45}\r$
must be different from zero. In addition, if $\l{\bar\Phi}_{1}^-\r$
and $\l\xi_2\r$ vanish the hidden matter mass matrix is singular and
the vacuum exhibits a runaway
behavior [\SQCD]. If we insist that the vacuum is well defined and that
the determinant of the matter condensates is nonvanishing, we see that
there is no set of VEVs that satisfies the new
set of F constraints, $\partial(W_3+W_5)/\partial\phi_i=0$, up to $N=5$.
Consequently, for all possible choices of VEVs,
$\partial W/\partial\phi_i\ne0$,
for some $\phi_i$. As long as $V_{eff}$ has a minimum, $\partial W/
\partial\phi_i\ne0$ in the minimum.

The above arguments apply to all orders $N>3$. In fact, we can
demonstrate this by considering the $N=7$ term, $W_7$, and its corrections to
the cubic level and $N=5$ F constraints. An analysis similar to the above for
the $N=5$
case shows that $\l W_7 \r \not=0$ if we require a stable vacuum. In addition,
the F equations for $\bar V_2, V_3, \Phi_{45}, \bar \Phi_{13},\xi_1$ are
modified due to $W_7$ because of the dependence of $\Pi_{32}$ on these
fields through the gaugino condensate and ${\cal M}^{-1}$. One can show that
now, these modified equations cannot be satisfied simultaneously with the ones
left unchanged. We conclude that $\l W \r \not=0$ and $\l W_i \r \not=0$ at
order $N=7$ too. Therefore, we expect that, generically $\l W \r \not=0$ and
$\l W_i \r \not=0$ (for some fields $\phi_i$) at some order $N>3$.

To show that supersymmetry is broken, we need to show that, at a given order
$N$, $\l F_i \r =exp(\l K \r/2)\l (WK_i+W_i) \r\not=0$.
We will show that $\l F_i \r$ obtained from the modified
superpotential, $W=W_3+\ldots+W_N$, is dominated by the $\l W_i \r$ piece.
(This unless $\l W \r$ arises from a lower order than $\l W_i \r$. We will
consider this case which is not the one at hand separately later.)
Then, whether $\l W \r$ vanishes or not at that order, the first term in
$\l F_i \r$ cannot cancel the second one.
In our case, $\l W_i \r \not=0$ for $i=\Phi_{45},\bar \Phi_1^-,\xi_2$ up to
order $N=5$. We expect that at higher orders there
will be other fields $\phi_i$ with $\l W_i \r \not=0$ as a result of
nonrenormalizable terms in the superpotential. (For example the $N=7$ terms
give in addition to the above fields $i=\bar V_2,V_3,\bar \Phi_{13}$.)
Assuming $\l\phi_i\r\approx M/10$, for matter with
$K=-c{\l\phi_i\phi_i^\dagger\r}/{M^2}$ and $c$ of order one
$$\l exp({K/2})\r\approx exp({-{\l\phi_i\phi_i^\dagger\r}\over{M^2}})\approx
exp({-{1\over{200}}})\sim1,\eqno(28)$$
where we took the dilaton and moduli VEVs to be of order $M$. Writing
the powers of $M$ explicitly,
$$\l WK_i\r\sim\l W_i\r
{{\l\phi_i\phi_i^\dagger\r}\over{M^2}}\approx 10^{-2}\l W_i\r.\eqno(29)$$
So the first term in $\l F_i\r$ is suppressed by a factor of
${\l\phi_i\phi_i^\dagger\r}/{M^2}\approx 10^{-2}$ with respect to the second
term at each order separately. The first term in $\l F_i \r$ can cancel
the second one either if it contains fields with VEVs larger than $M_{Pl}$
or if it arises from a lower order. We disregard the first possibility
because in this case the truncation of nonrenormalizable terms at any order
is inadequate.
{}From the relative magnitude
of the two terms ($\sim 100$) one sees that if $\l W \r$ arises from order
$N-2$ where $\l W_i \r$ arises from order $N$ the two terms in $\l F_i \r$
will be of the same
magnitude and they might cancel each other. This does not happen above, since
in our case both $\l W \r$ and $\l W_i \r$ become nonzero at the same order,
$N=5$.
In general, such a cancellation requires a high degree of fine
tuning of the scalar VEVs and may even be excluded due to the other F and D
constraints.

In our case, both $\l W \r$ and $\l W_i \r$ become nonzero at $N=5$ and
the $W_i$ piece (from $N=5$) dominates $\l F_i \r$.
Consequently, $\l F_i\r\sim\l W_i\r\Rightarrow\l F_i\r\ne0$ if $\l W_i\r\ne0$
and supersymmetry is broken by the $\phi_i$ (matter) F term.
We have argued above that $\l F_i \r \not=0$ for all sets of VEVs or minima of
$V_{eff}$. This means that supersymmetry cannot be preserved by any of the
vacua of the new, modified superpotential. As long as $V_{eff}$ has a minimum,
supersymmetry will be broken in that minimum. Although we cannot show
its existence explicitly, we assume that $V_{eff}$ has at least one minimum
(which is the new vacuum).

For simplicity we considered the corrections to $W_3$ only up to $N=5$.
However, our arguments are valid to all orders $N$.
In fact, if for some cubic level flat direction the nonrenormalizable terms up
to order $N=5$ vanish, the same mechanism will break supersymmetry at some
higher order (e.g. $N=7$ in our case).
We expect that once supersymmetry is broken at some order $N$,
terms from higher orders cannot restore it. The reason is that higher order
terms need to include scalars with VEVs larger than the Planck scale $M_{Pl}$.
For example, since $N=7$ order terms
are suppressed by $10^{-2}$ with respect to $N=5$ order terms,
to cancel $\l (W_3+W_5)_i \r$ from $N=7$ terms we need, for some
scalars which appear in the $N=7$ order terms,
$\l \phi \r \sim 10^2 (M/10) \sim 10M \sim M_{Pl}$. However, generically
the VEVs are of order $g^2M/4\pi\sim M/10$. The situation is
complicated in the case with matter condensates because of the $\Pi_{ij}$
dependence on ${\cal M}_{ij}^{-1}$. However, even if, due to
the matter condensates, some
term happens to be of the same magnitude as some lower order term, without
a high degree of fine tuning, we again expect that if
supersymmetry is broken at some order $N$, it is not restored by higher
orders.

To estimate the magnitude of supersymmetry breaking in the observable
sector, we estimate the gravitino mass which is given by
$$m_{3/2}={1\over{M^2}}e^{\l{K}/2\r}\vert\l{W}\r\vert,\eqno(31)$$
With $\Lambda_5\sim10^{13}~GeV$ the gaugino condensate is estimated from
Eq. (19a) to be $\sim10^{39}~GeV^3$.
For the matter condensates we get from Eq. (19b) and Eq. (22),
$$\Pi_{11}\approx10^{24}~GeV^2{\hskip 1cm}\Pi_{23}
\approx10^{26}~GeV^2{\hskip 1cm}\Pi_{32}
\approx10^{27}~GeV^2\eqno(32)$$
Taking $\l\phi\r/M\sim1/10$ we get $m_{3/2}\sim 1~TeV$ from the $N=5$,
$m_{3/2}\sim 10~TeV$ from $N=7$ order terms and
$m_{3/2}\sim 100~GeV$ from the $N=8$ order terms.
Using the modular invariant expressions for the gauge and matter
condensates from Ref. [\LT] and multiplying the nonrenormalizable terms
by appropriate factors of the Dedekind $\eta$ function, as outlined in Ref.
[\KLN], yields results that differ from ours at most by an order of magnitude
which is tolerable for our qualitative estimates.

\bigskip
\vfill
\eject
\centerline{\bf 5. Discussion and conclusions}

In this paper, we examined the problem of
supersymmetry breaking in a class of realistic
superstring derived standard--like models.
We presented a possible scenario for supersymmetry breaking that takes
into account modifications of the cubic level superpotential
due to nonrenormalizable terms and hidden sector matter condensates.
Hidden matter condensates form when the hidden gauge group becomes strong at
a scale, $\Lambda_H \sim 10^{12-14}~GeV$. At the same scale, gaugino
condensates form as well.
Supersymmetry breaking due to gaugino condensation
has been extensively studied in the literature [\GSUSYB].
Is it possible to say under
what conditions either of the two supersymmetry breaking mechanisms is
dominant? Supersymmetry is broken when
$$\l F_i\r=\l e^G (G_i^j)^{-1}G_j\r +\l f_{\alpha \beta,i} \lambda^{\alpha}
\lambda^{\beta}\r\ne0~. \eqno(33)$$
At the string tree level $f_{\alpha \beta}=S \delta_{\alpha \beta}$ and
the second term is nonzero only for $i=S$. We want to estimate the
relative magnitudes of these two contributions. The first and the second
terms in $\l F_i \r$ arise from, scalar matter condensation
and from gaugino condensation, respectively.
The values
of the gaugino and matter condensates are given by Eqs. (19a,b).
Note that $\Pi_{ij}$ and
therefore $F_i$ from matter condensation are proportional to the gaugino
condensate $\l \lambda^{\alpha} \lambda^{\beta} \r$. Taking generically,
$\l \phi \r \sim M/10$ and using the fact that the mass matrix $\cal M$
arises from order $N>3$ terms, we get from nonrenormalizable terms of order
$L$,
$$W_L=\Pi_{ij}{{\l\phi\r^{L-2}}\over{M^{L-3}}}~~~\hbox{and}
{}~~~\Pi_{ij}\sim{\l \lambda^{\alpha} \lambda^{\beta}\r\over {32\pi^2}}
{{M^{N-3}}\over{\l\phi\r^{N-2}}}~. \eqno(34)$$
Therefore,
$$\l F_\phi \r \sim {\l \lambda^{\alpha} \lambda^{\beta}\r\over {32\pi^2}}
\left({{\l\phi\r}\over{M}}\right)^{L-N}
{1\over{\l\phi\r}}~. \eqno(35)$$
The contribution of the gaugino condensate to the F--term is simply given by
$\l F_g\r\sim{\l\lambda^{\alpha}\lambda^{\beta} \r/ {32\pi^2M}}$.
We expect, by comparing the two expressions, that matter
condensation effects are dominant when $L-N<1$ and
vice versa.
In general, whether one or the other
supersymmetry breaking
mechanisms is dominant depends not only on the hidden gauge group that
condenses and its matter content but also on the orders at which hidden matter
gets mass and F equations are violated as given by the relation $L-N<1$. In
this respect our model is on the borderline, i.e. the contributions of hidden
matter and gaugino condensation to supersymmetry breaking are comparable since
in our case $L \sim N$.

A fundamental problem in all
supersymmetry breaking mechanisms is the stability
of the dilaton vacuum.
Perturbative coupling unification requires that
$g^2(M_{Pl}) \sim 1/4\l S \r \sim 1/2$. We have not calculated $V_{eff}$
explicitly and therefore cannot say whether or not in our case the dilaton
potential
is stable with the required minimum. It has been previously noted that
one can stabilize the dilaton potential if there are two hidden gauge groups
which condense. The superstring standard--like models typically admit
two non--Abelian hidden gauge groups, $SU(5)\times SU(3)$. Of those
one has to be broken at $M$ by VEVs that are needed to obtain
realistic quark mixing and a stable vacuum. In the scenario that we
considered the hidden $SU(3)$ is broken by these VEVs and therefore dilaton
stabilization by two (or more) non--Abelian hidden gauge groups cannot be
applied. However, there exists a possibility to break the hidden $SU(5)$ gauge
group, for example to $SU(3)\times SU(2)\times U(1)$ or to
$SU(4)\times U(1)$, in which case it may be possible to
obtain two hidden gauge groups
with matter content and similar beta functions. In this case we
may have a problem with vacua for which the number of, $N\oplus {\bar N}$
flavor pairs exceeds, $N$, the number of colors. Recent progress on
nonperturbative supersymmetric field theories may be instrumental in this
case [\S]. Another possibility is to try
to build the effective superpotential in terms of the dilaton, $S$
and the moduli, $\hat T_i$. For example, the nonrenormalizable terms produce a
moduli dependence from the
Dedekind $\eta$ functions which
are included to get modular weight $-3$.
The dilaton dependence arises from $\Lambda \sim exp(-8\pi^2 S/b)$,
which appears in $\Pi_{ij}$, and from
$\l \phi \r \sim g^2M^2/4\pi \sim M^2/4\pi S$.
One can try to build these effective terms and minimize them for $S$ and
$\hat T$.

Another fundamental problem is the value of the cosmological constant
($\Lambda_C$) after
supersymmetry breaking. The condition for a vanishing cosmological constant
is, from Eq. (23),
$$\l (G_i^j)^{-1} G_j G^i \r=3~, \eqno(36)$$
where a summation over $i,j$ is implied. This is an additional constraint
that the supersymmetry breaking vacuum must satisfy. The problem with this
constraint is that an approximate (order by order in $N$) calculation is
meaningless since a negligible
effect of order $m/M$ (where $m<<M$) results in $\Lambda_C \sim m^2 M^2$
which is still huge.
We expect that in our model $\Lambda_C \not=0$ and we do not have much to
add on this point.

A novel feature of free fermionic models is
the existence of horizontal gauge $U(1)_r$ symmetries under which both the
observable and hidden matter states which eventually condense are charged.
Therefore, supersymmetry breaking in the hidden sector can be communicated to
the observable sector not only by gravity (which is universal) but also by
different $U(1)_r$ which may couple the observable and hidden sectors
nonuniversally. This coupling will depend on the $U(1)_r$ charges of the
relevant states and the mass of the $U(1)_r$ gauge boson. Both of these can
be slightly different for different observable matter states. Thus, in these
models one may obtain different soft  supersymmetry breaking masses for the
scalars.

The low--energy phenomenology which arises from supersymmetry breaking due to
nonrenormalizable terms and
hidden matter condensation is particularly interesting.
In general, the gaugino and matter condensates generate nonvanishing
F--terms for the dilaton, the moduli and the matter fields. The
various F--terms produce different forms of soft supersymmetry
breaking terms. The dominant supersymmetry breaking F--term will determine
the supersymmetric mass spectrum and consequently the low energy
phenomenology. For example, if the dilaton F--term dominates, squark masses
are universal [\KL], while if the moduli terms dominate, in general,
one expects the squark masses to be nonuniversal [\IL]. For specific
cubic level F and D flat solutions, like Eq. (14), it may be possible,
as we argued above, to determine the dominant term and therefore
to make specific predictions for the supersymmetric mass spectrum.

In this paper we examined the problem supersymmetry breaking
in the superstring derived standard--like models. The problem
of supersymmetry breaking has two different aspects. The
first is the determination of the compactification parameters,
i.e. the dilaton, the moduli and the remaining $SO(10)$ singlet VEVs.
The determination of this parameters must await a better understanding
of superstring theory both at the perturbative as well as the
nonperturbative level. It may be futile, in our opinion, to try to
determine these parameters only by incorporating nonperturbative effects
in the effective point field theory. However, if we accept this deficiency,
while seeking a better understanding of string theory we can
parameterize our ignorance into several parameters like the dilaton VEV,
the moduli VEVs and the singlet VEVs. We showed that with these parameters
fixed supersymmetry breaking arises due to the existence of a non--Abelian
hidden gauge group with matter in vector--like representations.
The modification of the superpotential due to the strong hidden
sector dynamics and nonrenormalizable terms may result in hierarchical
supersymmetry breaking in the observable sector.
The resulting soft supersymmetry breaking terms can then be obtained
and specific predictions for the supersymmetric spectrum be made.
 The existence of
supersymmetric particles near the TeV scale will be decided by future
experiments. In the event that supersymmetric particles are observed
the sparticle spectrum will be used to restrict the SUSY breaking
parameters and hence will be used to study the compactification
parameters and their determination by possibly nonperturbative
string effects. We will return to the pursuit of these ideas
in future publications.

\bigskip
\centerline{\bf Acknowledgments}

We thank Jan Louis for useful discussion.
AF was supported by an SSC fellowship.
EH was supported by the Department of Particle Physics and a Feinberg
Fellowship.

\vfill
\eject
\refout
\end